\documentclass[aps,11pt,notitlepage]{revtex4-1}

\usepackage{graphicx}  
\usepackage{dcolumn}   
\usepackage{bm}        
\usepackage{amssymb}   
\usepackage{amsmath}

\newcommand{\1}{$|1\rangle$}
\newcommand{\2}{$|2\rangle$}

\begin{document}

\title{High momentum splitting of matter-waves by an atom chip field gradient beam-splitter}
\author{Shimon Machluf, Yonathan Japha and Ron Folman}
	\email{folman@bgu.ac.il}
	\affiliation{Department of Physics, Ben-Gurion University, Be'er Sheva 84105, Israel}
\date{\today}

\begin{abstract}
The splitting of matter-waves into a superposition of spatially separated states is a fundamental tool for studying the basic tenets of quantum mechanics and other theories, as well as a building block for numerous technological applications. We report the realization of a matter-wave beam splitter based on magnetic field gradients on an atom chip, which can be used for freely propagating or trapped atoms in a Bose-Einstein condensate or a thermal state. It has a wide dynamic range of momentum transfer and operation time. Differential velocities exceeding 0.5 m/s can be achieved in a few micro-seconds. The beam splitter may enable a wide range of applications, such as, fundamental studies of many-body entanglement and dephasing processes, probing classical and quantum properties of nearby solids, and metrology of rotation, acceleration and gravity on a chip scale.
\end{abstract}

\maketitle

\section{Introduction}

For the last two decades, matter-waves, in the form of ultra cold atoms \cite{CCTNobel, WKNobel}, have been the source of numerous new experimental insights. One of the basic tools for such studies is matter-wave interferometry which enables high precision measurements for fundamental physics \cite{old41,old43,old1, old8, old7, fund1} as well as for technological applications \cite{kasevich-rotation,old6,app1,app2,app3}.

In one type of matter-wave interferometry, atoms are split into a superposition of momentum states by interaction with laser light, where the signal is typically derived from the population of internal states \cite{kasevich,HighMomentum1,HighMomentum2,HighMomentum3,HighMomentum4}. High precision measurements, e.g. of acceleration and rotation, are enabled by an accurate momentum transfer from laser photons to atoms in multiples of the photon momentum $\hbar k$, where $k=2\pi/\lambda$ is the wave-vector of the photons.

In another type of interference experiments, a Bose-Einstein condensate (BEC) is split into two parts by a potential formed by optical \cite{optical,cornell}, radio-frequency (RF) or micro-wave \cite{joerg,treutlein} fields. The signal is typically derived from spatial interference patterns. These experiments provide new experimental insight in topics such as coherence and entanglement in a many-body system \cite{Esteve, 1DBEC, yoni2}.

Potential applications of spatial interferometry with a BEC on an atom chip \cite{RonRev2002,ReichelRev,fortagh}, which is the setting used here, include the study of properties of many body systems, as well as the sensitive probing of classical or quantum properties of solid state nano-scale devices and surface physics (e.g. \cite{SQUID1,SQUID2, shotNoise1,shotNoise2,milton}). The latter is expected to enhance considerably the power of non-interferometric measurements with ultracold atoms on a chip, which have already contributed, for example, to the study of long-range order of current fluctuations in thin films \cite{OurScience}, the Casimir-Polder force \cite{Vuletic,Ketterle1,Ketterle2,cornellCP} and Johnson noise from a surface \cite{Jones}.

In addition, coherent splitting of matter-waves on a chip is a crucial step towards the development of miniature rotation, acceleration and gravitational sensors based on guided matter-waves \cite{yoni,Baker}.
Depending on the progress made in developing stable electronics, a current-carrying atom chip may be used for metrology with either guided or freely propagating atoms.

Here we propose an atom chip field-gradient-beam-splitter (FGBS), which may provide an efficient and flexible tool for interferometry in the micrometer scale as well as a macroscopic scale. The operation of the FGBS is simple and in the present realization requires only the manipulation of static magnetic and RF fields. The FGBS enables a large momentum transfer in a short time, and as the coherence time of split matter-waves is limited by atom-atom interaction and coupling to the environment, it may enable sensitive interferometry, for example by way of large density and large area, within the available coherence time. We present a specific realization of the FGBS using magnetically sensitive Zeeman levels and demonstrate the formation of spatial interference fringes as an interferometric signal for both freely propagating and trapped atoms. We also propose other realizations using first-order magnetically insensitive states ("clock states") and internal state population as a signal.

\section{General scheme}

\begin{figure}[!t]
\centering
\includegraphics[width=\textwidth]{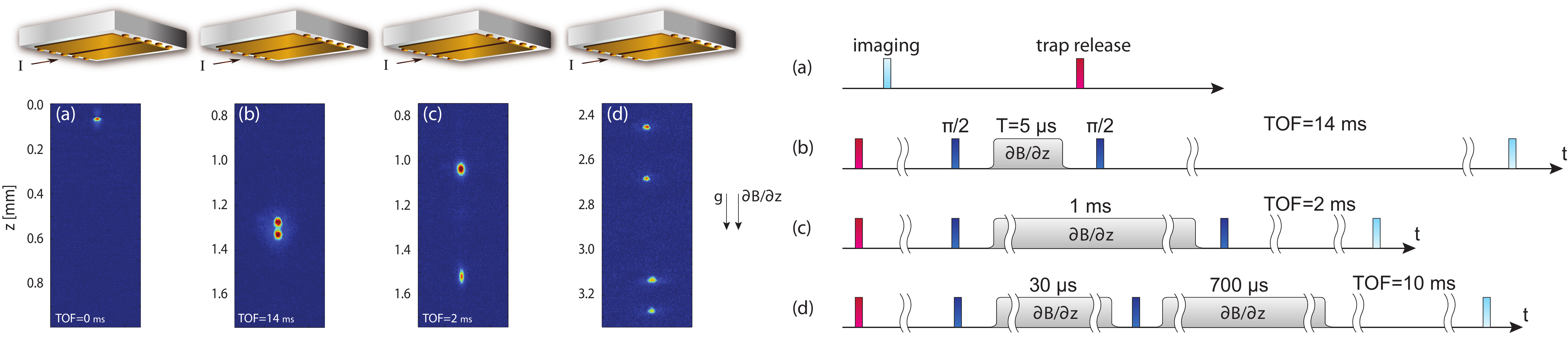}%
\caption{
Absorption images of the atoms and the corresponding experimental sequence: (a) in the trap before the release, (b) after weak splitting of less than $\hbar k$ in $5\,\mu$s interaction time and $14\,$ms time-of-flight (TOF), (c) after a strong splitting of above $40\,\hbar k$ in $1\,$ms interaction time and $2\,$ms TOF, and  (d) the four clouds separated by another strong gradient (Stern-Gerlach) after the FGBS. Images (b) and (c) show the high dynamic range this method can reach without any complicated sequence. Image (d) shows that each cloud in (b) and (c) is constructed from atoms in the $|1\rangle$ and $|2\rangle$ states (see text for details).
}%
\label{fig:FGBS:fig1}%
\end{figure}

The general scheme of the FGBS and its output is demonstrated in Fig. \ref{fig:FGBS:fig1}. It is based on a Ramsey-like sequence of
two $\pi/2$ rotations with a field gradient applied during the interval between them.
If the gradient is applied perpendicular to the motion of the atoms, the differential momentum will be in the transverse direction and an area enclosing interferometry would ensue.
In our experimental demonstration the gradient is applied along the direction of motion to form a one dimensional interferometer.
Considering two-level atoms with internal states $|1\rangle$ and $|2\rangle$ and starting with atoms in state $|2\rangle$, the first $\pi/2$ rotation transfers them into the superposition state $\frac{1}{\sqrt{2}}(|1\rangle+|2\rangle)$.
We then apply a field gradient which constitutes a state-selective
force $F_j(j=1,2) = -\nabla{ V_{j}}$ for an interaction time $T$, which is typically shorter than the time it takes the atoms to move in
the force field. The potential $V$ may be due to any kind of field such as electro-magnetic, electric or magnetic, and the states \1 and \2 may be any states between which one can introduce a coupling and for which there exists a state selective force. The state of the atoms after time $T$ is given by $|\psi({\bf x},T)\rangle=
\frac{1}{\sqrt{2}}\left(|1\rangle e^{i{\bf F}_1\cdot{\bf x}T/\hbar}+|2\rangle e^{i{\bf F}_2\cdot{\bf x}T/\hbar}\right)\psi_0({\bf x})$, where
$\psi_0({\bf x})$ is the spatial wave-function at $t=0$. Each level
gains a phase $\phi({\bf x},t)=-\Delta E_j t/\hbar={\bf F}_j\cdot{\bf x}T/\hbar$, which is equivalent to a momentum transfer
${\bf p}_j={\bf F}_j T$.
The second $\pi/2$ rotation transfers the atoms into the superposition state
\begin{equation}
|\psi\rangle=\frac{1}{2}\left(|1,{\bf p}_1\rangle+|1,{\bf p}_2\rangle-|2,{\bf p}_1\rangle+|2,{\bf p}_2\rangle\right)
\label{eq:FGBS:general}
\end{equation}
such that each of the internal states $|j\rangle$ is in a superposition of two momentum states $|{\bf p}_j\rangle\equiv e^{i{\bf p}_j\cdot{\bf x}/\hbar}\psi_0({\bf x})$, and vice versa. Eq.~(1) is also valid if the atomic motion during the interaction time is taken into account or when the force is not homogeneous in space and time, given that the momentum states $|{\bf p}_j\rangle$ are more generally defined as the states that evolve from $\psi_0({\bf x})$ under the influence of the respective potentials $V_j({\bf x},t)$.

This completes the operation of spatial splitting.
One may then use another field gradient pulse to separate between the \1 and \2 states [Fig.  \ref{fig:FGBS:fig1}(d)], and use these states to realize two parallel interferometers for noise rejection.
One can also use the entangled momentum and internal state as an interferometer of two clocks (e.g. \cite{nature_comm}).
If one wishes to stay with just one of the two internal states, one may typically find a dedicated transition to discard the redundant state.

\section{Experiment and results}

To realize the FGBS, we utilize Zeeman sub-levels.
The idea of splitting different spin states in a magnetic gradient is as old as the Stern-Gerlach experiments of the 1920s (e.g. \cite{SGApp}), but to the best of our knowledge was never designed with the necessary detail and to this day remained unrealized.
Furthermore, unlike such schemes, the FGBS gives rise at its output to two wavepackets with identical spin states, an advantageous feature in noisy environments.

\subsection{Splitting free falling atoms}

We start with a BEC of $\sim10^{4}\ ^{87}Rb$ atoms in state $|F,m_F\rangle=|2,2\rangle\equiv |2\rangle$ and use an RF field to perform transitions to state $|1\rangle \equiv |2,1\rangle$.
We utilize the same set-up described in \cite{shimi}. The trap position is 100$\,\mu$m from the chip surface, and the radial trapping frequency of \2 is  $\omega_{2}\approx2\pi\times100\,$Hz (for state \1, $\omega_{1}=\omega_{2}/\sqrt{2}$).
In order to have the $|1\rangle$ and $|2\rangle$ states form a pure two-level system, we apply a strong homogeneous magnetic field ($\Delta E_{12}\approx h\times25\,$MHz) and push the transition to $|2,0\rangle$ out of resonance, due to the nonlinear Zeeman effect ($\sim250\,$kHz) \cite{shimi}. 
Next, we release the BEC and apply two $\pi/2$ RF pulses  \cite{cataliotti} with a Rabi frequency of $\Omega_{R}=20-25\,$kHz and with a magnetic gradient pulse of length $T$ in between, thus forming a Ramsey-like interferometer.

This scheme for freely falling atoms has been applied for atoms above and below the BEC transition temperature.
Here we present our results for a BEC.
Fig.~\ref{fig:FGBS:fig1}(a) presents the cloud position in the trap, before it was released.
In Fig. \ref{fig:FGBS:fig1}(b,c) we present two splitting events which exhibit the large dynamic range of this method,
one with a differential momentum of less than $\hbar k$ ($T=5\,\mu$s), and the other with above $40\,\hbar k$ ($T=1\,$ms).
Here $\hbar k$ is a reference momentum of a photon with $1\,\mu$m wavelength.
The only parameter which is changed is the interaction time $T$, in which  a current of $2-3\,$A in a $200\times2\,\mu$m$^{2}$ gold wire on the chip surface gives rise to the gradient.
To verify the internal state of the atoms [Eq. \eqref{eq:FGBS:general}] we separate between the \1 and \2 states by a long pulse of magnetic field gradient (Stern-Gerlach), as shown in Fig.~\ref{fig:FGBS:fig1}(d).

\begin{figure}[t]%
\centering
\includegraphics[width=0.75\textwidth]{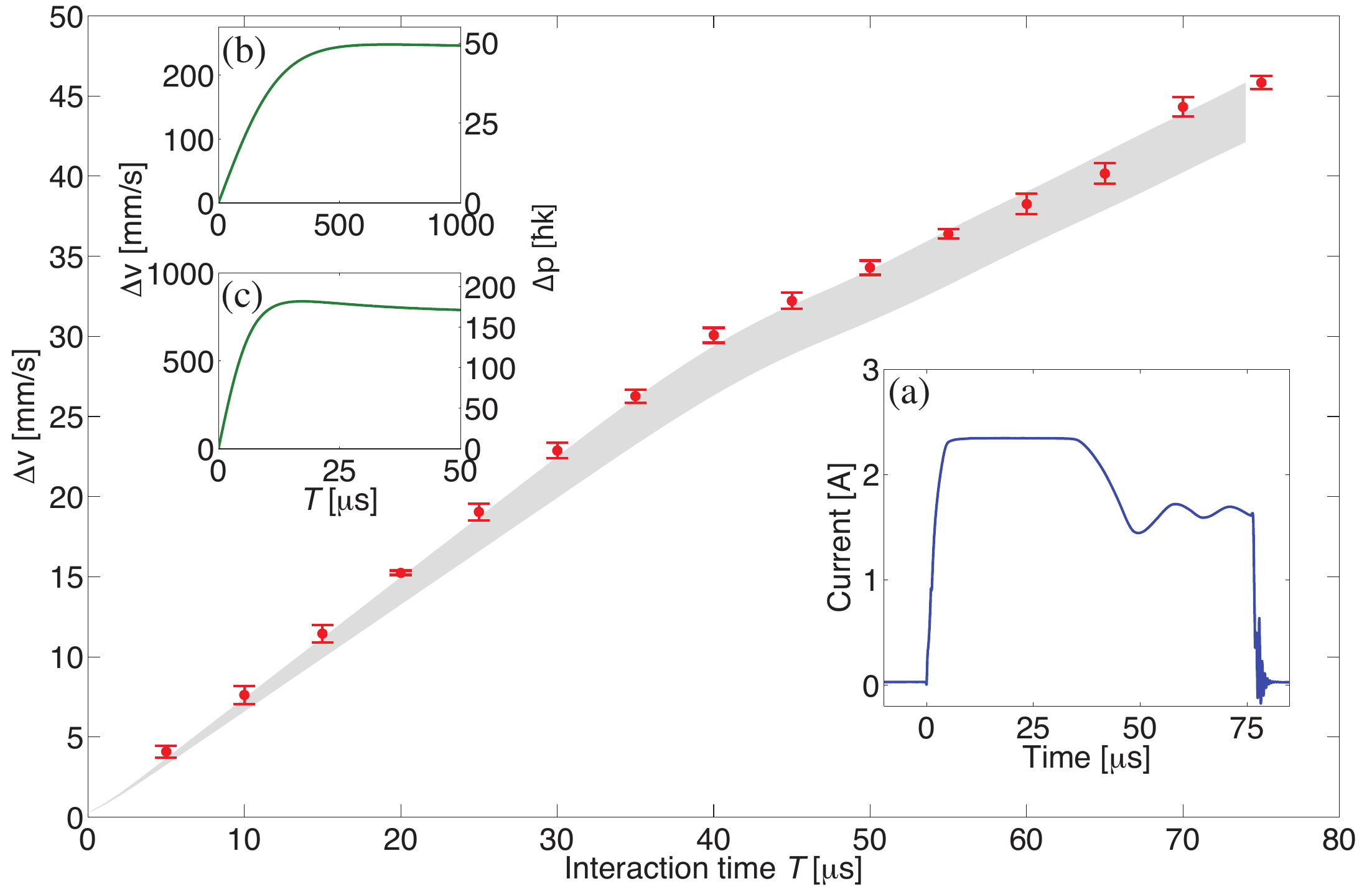}%
\caption{
Characterization of the FGBS for free falling atoms: The differential velocity between the two observed clouds as a function of the interaction time $T$.
The error bars are calculated from the variation of a few data sets.
The theoretical prediction (shaded area) is based on a measured current in a resistor that mimics the chip wire [inset (a)], taking into account errors of $\pm 2.5\,\mu$m (one pixel) in the initial cloud position, $\pm 0.2\,\Omega$ in the wire resistance and a 1$\,\mu$s delay of the measured rise-time (due to the resistor's inductance).
The linear dependence on $T$ is to be expected from the solution of Eq.~(\protect\ref{eq:FGBS:dpgrad}). The current ``overshoot'' at short times is responsible for the larger acceleration for small $T$.
Inset (b): Numerical integration of Eq.~(\protect\ref{eq:FGBS:dpgrad}) over  $T$, using the experimental wire configuration, and using a constant current of $3\,$A.
We observe the limit of our method due to the large distance which develops between the atoms and the wire.
Inset (c): Utilizing improved experimental parameters we find that $\Delta p$ of above $100\,\hbar k$ is possible in less than $10\,\mu$s.
The parameters are: $z=10\,\mu$m, $I=2\,$A and  the wire dimensions are $10\times2\,\mu$m$^{2}$ ($10^7\,$A/cm is more than sustainable for such short pulses).
}
\label{fig:FGBS:fig2}%
\end{figure}

Fig.~\ref{fig:FGBS:fig2} shows the measured differential momentum transfer as a function of the interaction time $T$. An interaction time as short as $100\,\mu$s is required to transfer a relative velocity of $50\,$mm/s (equivalent to $10\,\hbar k$). Continuing the general theoretical description given above, the operation of the magnetic realization of the FGBS may be explained by simple kinematics. During the interaction time $T$, a differential acceleration between the wavepackets is induced, such that after the FGBS, each internal state is a superposition of two wavepackets which were accelerated as a \1 or \2 state. The momentum kick for a wavepacket of a certain $m_F$ state at a distance $z$ below the chip wire is
\begin{equation}
\left.\frac{dp}{dt}\right|_{m_F}=\frac{m_F g_F\mu_B\mu_0 I }{2\pi z^2}
\label{eq:FGBS:dpgrad}
\end{equation}
where $\mu_0$ and $\mu_B$ are the magnetic permeability of free space and the Bohr magneton, $g_F$ is the Land\'e factor for the hyperfine state $F$, and $I$ is the current. The equation does not present the nonlinear term in $B$ and a geometric term $1/[1+(W/2z)^2]$, accounting for the finite width $W$ of the wire.
These terms have not been
neglected in our full simulation of the FGBS (in Fig. \ref{fig:FGBS:fig2}) which predicts the experimental results and shows that momentum transfers of over 100$\,\hbar k$ are feasible in a few micro-seconds
[see inset (c), Fig. \ref{fig:FGBS:fig2}].
The linear relation in Fig. \ref{fig:FGBS:fig2} is to be expected for the short interaction times during which the atoms move only slightly and the acceleration in Eq. \eqref{eq:FGBS:dpgrad} is fairly constant (the kink in Fig. \ref{fig:FGBS:fig2} is due to changing currents, see caption).

\begin{figure}[!t]
\centering
\includegraphics[width=0.65\textwidth]{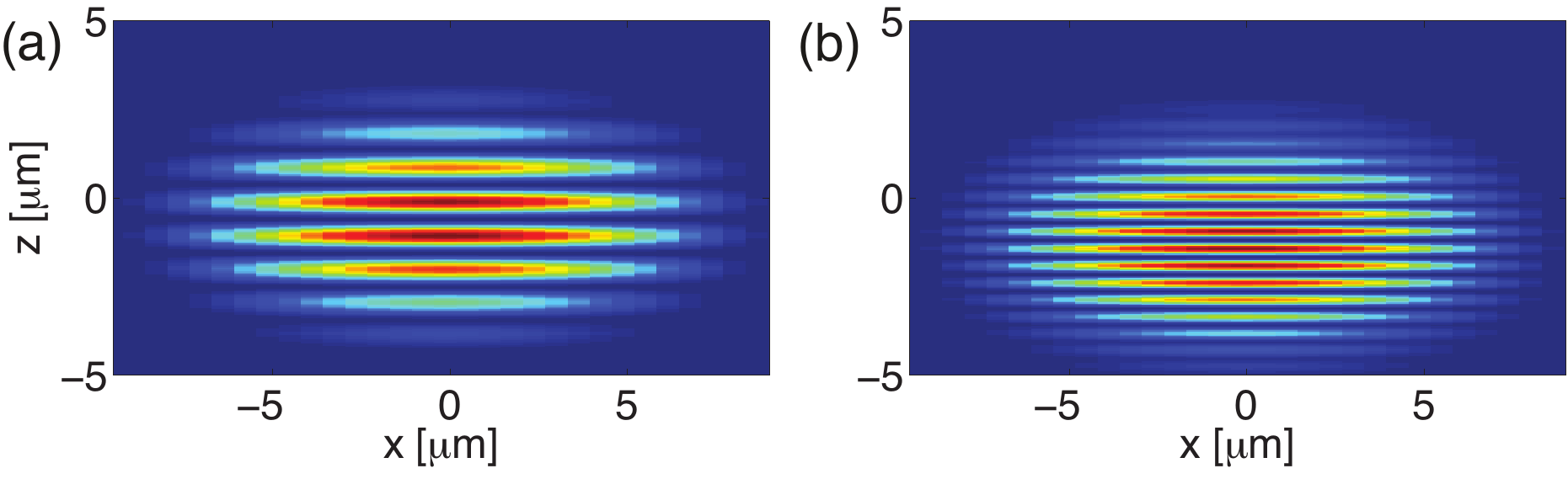}
\caption{Simulation of the free fall experiment: For (a) $T=5\,\mu$s and (b) $T=10\,\mu$s, we present the atomic density of the \2 state (GP simulation), just after the splitting. These ``near field'' fringes are Fourier transformed into two wavepackets in the ``far field''. This point of view is an alternative to the simple kinematic explanation provided by Eq.~(\protect\ref{eq:FGBS:dpgrad}).}
\label{fig:FGBS:insituFringes}
\end{figure}

In Fig. \ref{fig:FGBS:insituFringes} we present a Gross-Pitaevskii (GP) simulation of the atomic density for state \2 just after the second $\pi/2$ pulse (``near field").
The density shows a sinusoidal pattern.
This pattern may be viewed as spatial Ramsey fringes: an atom at  position ${\bf x}$ is subjected to a Ramsey sequence in which the phase accumulated between the two $\pi/2$ pulses is $ 1 /\hbar \int_{0}^{T} \Delta V({\bf x}, t)\,dt$ where $\Delta V$ is the potential difference and $T$ is the interaction time.
As the potential varies linearly along the ${\hat z}$ axis, $\Delta V = \Delta F z$, and at short times the atom stays almost stationary, the Ramsey phase and consequently the population of the two states \1 and \2 is modulated with a wavelength $\lambda=h/\Delta F T$. The formation of sinusoidally modulated internal state populations is equivalent to varying Rabi frequencies along the ensemble when the driving field has an intensity gradient \cite{Ruti,Treutlein1,Anat}, when $\Delta F$ is replaced by the gradient of the Rabi frequency (up to $\hbar$).
As a perfect sine function Fourier transforms into two identical and counter propagating $k$ components, this ``near field" density distribution transforms after  a sufficient time-of-flight into a pair of wavepackets with momentum difference $\Delta p=h/\lambda$.
We note that this is the inverse of a recombination process of two wavepackets.

\subsection{Splitting trapped atoms}

\begin{figure}[!t]
\centering
\includegraphics[width=\textwidth]{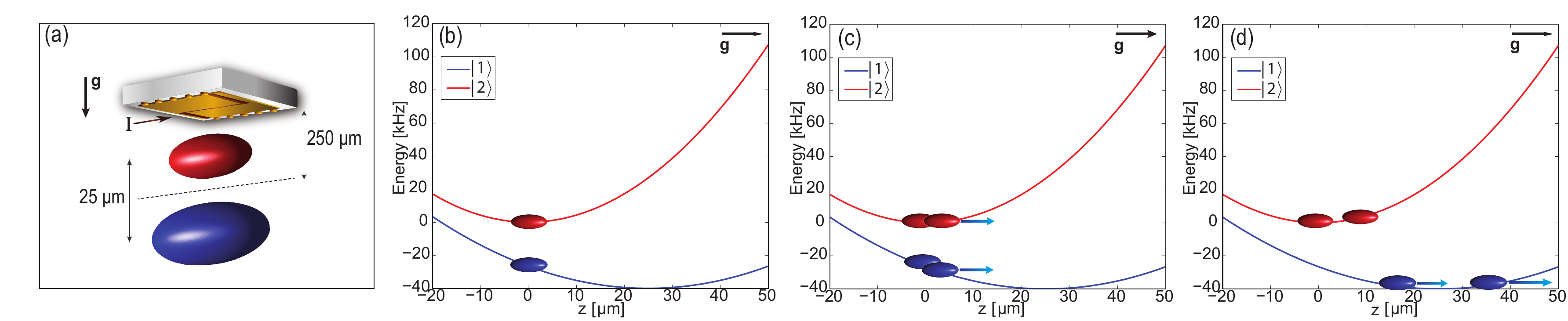}%
\caption{
Splitting a trapped BEC: (a) An illustration of the two traps below the atom chip showing the vertical separation between the two traps, and the difference in size due to the different confinement.
(b-d) Trapping potential of the two states (the energy difference was minimized for visibility) in the direction of gravity during the splitting process.
The four parts of the wave-function and their momentum (size of arrow) are also shown.
(b) The splitting just after the first $\pi/2$ pulse, where the position of both clouds is the trap minimum of the \2 state.
(c) The clouds after the second $\pi/2$ pulse where only the \1 part of the super-position was accelerated before the pulse. There is still spatial overlap (the distance in exaggerated in the image for clarity).
(d) The four parts after some oscillation time in the trap, showing that one part of the \2 state didn't move, the other part of the \2 was slowed by the trapping potential almost to a halt, while the two parts of \1 were accelerated.
}%
\label{fig:FGBS:fig4}%
\end{figure}

Next, we show that our scheme could also be implemented on a trapped BEC, thereby opening the road for guided ``multi-pass" interferometry, such as the Sagnac we have previously proposed \cite{yoni}.
We repeat the above scheme and performing all the sequence on an atom cloud trapped in an Ioffe-Pritchard (IP) magnetic trap.
We start with $\sim10^{4}$ $^{87}Rb$ atoms in a trap $250\,\mu$m from the chip, with a radial trapping frequency of $\omega_{2}\approx2\pi\times100\,$Hz. 
Due to the high magnetic field at the trap bottom, $\Delta E_{12}\approx h\times18\,$MHz and the nonlinear Zeeman shift is about $100\,$kHz.
The sequence is similar to the free falling experiment but with the release taking place after the second $\pi/2$ pulse.
Contrary to the free falling case, no external gradient is required as the trap gradient naturally accelerates the atoms.
Furthermore, the gradient exists also during the $\pi/2$ pulses, but due to the wide spectrum of the short pulse it is still quite effective in creating the superposition (a $\pi$ pulse with $\Omega_{R}=5-10\,$kHz transfers up to $90\%$ of the atoms, compared with more than $95\%$ in free fall).

\begin{figure}[!t]%
\centering
\includegraphics[width=0.75\textwidth]{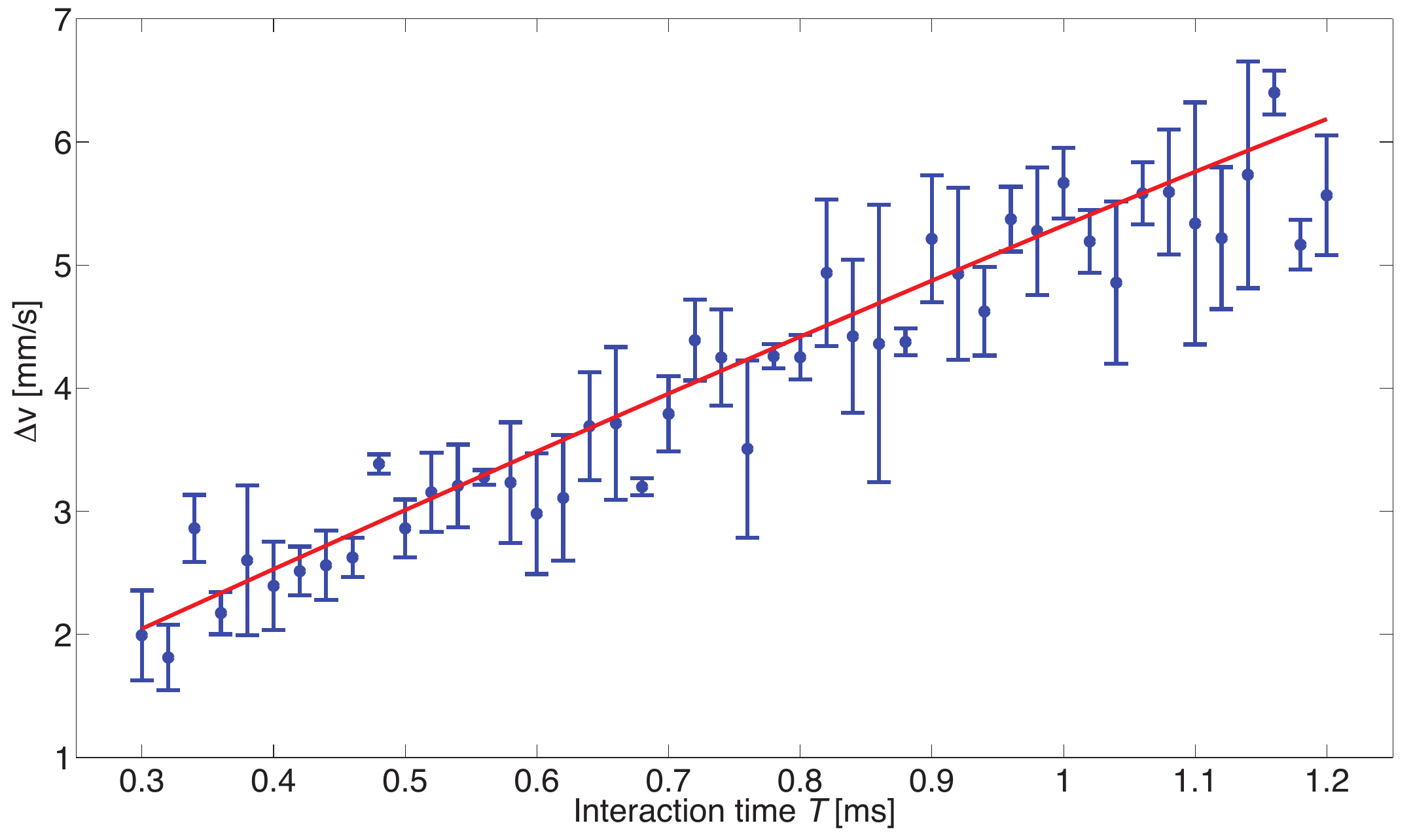}%
\caption{
Characterization of the FGBS for trapped atoms:
The differential velocity of the two wavepackets as a function of the interaction time $T$.
$\Delta v$, and its error, were taken from a linear fit of
six measurements for each $T$, at two different Rabi frequencies and three TOF.
As expected from Eq. \eqref{eq:FGBS:TrappedAcceleration}, $\Delta v$ grows linearly with time.
The solid line is a theoretical curve $v(T)=\frac{g}{2}\sin(\omega_{1} T)/\omega_{1}+v_r$, where  $v_r=0.58\,$mm/s is an additional velocity due to atom-atom repulsive interaction (no fitting parameters).
The first term follows from Eq. \eqref{eq:FGBS:TrappedAcceleration}, while the collisional constant $v_r$ is obtained from a full GP simulation.
The linearity of the graph is due to the small interaction times such that $\sin(\omega T)/\omega\approx T$.
}%
\label{fig:FGBS:fig5}%
\end{figure}

In Fig. \ref{fig:FGBS:fig4}(a) we present the two potentials by their equipotential surfaces, as they are situated below the atom chip. In Fig. \ref{fig:FGBS:fig4}(b-d) we utilize a 1D energy versus position ($\hat{z}$) plot to describe the evolution of the system during and after the FGBS sequence. Due to gravity, the centers of the combined magnetic and gravitational trapping potentials for the two levels are shifted by $\Delta z=g/{\omega_2}^2$.
It follows that when atoms initially at the level \2 are transferred by the first $\pi/2$ pulse into level \1, they experience an acceleration
\begin{equation}\label{eq:FGBS:TrappedAcceleration}
a(T)={\omega_1}^2 z(T)= \omega_{1}^{2}\Delta z \cos(\omega_{1}T) =\frac g 2 \cos(\omega_{1}T).
\end{equation}
For interaction times $T\ll \pi/\omega_1$ these atoms move only sightly along the potential gradient such that the acceleration is constant and equals $g/2$, 
and as in the free fall scheme the momentum splitting grows almost linearly with $T$. This almost linear dependence is shown in Fig. \ref{fig:FGBS:fig5}.
Here, the atom-atom collisional repulsion is responsible for an additional contribution to the velocity, as shown in the experimental data and confirmed
by our numerical GP solution.

\subsection{Recombining the two wavepackets}

\begin{figure}[!t]%
\centering
\includegraphics[width=0.75\textwidth]{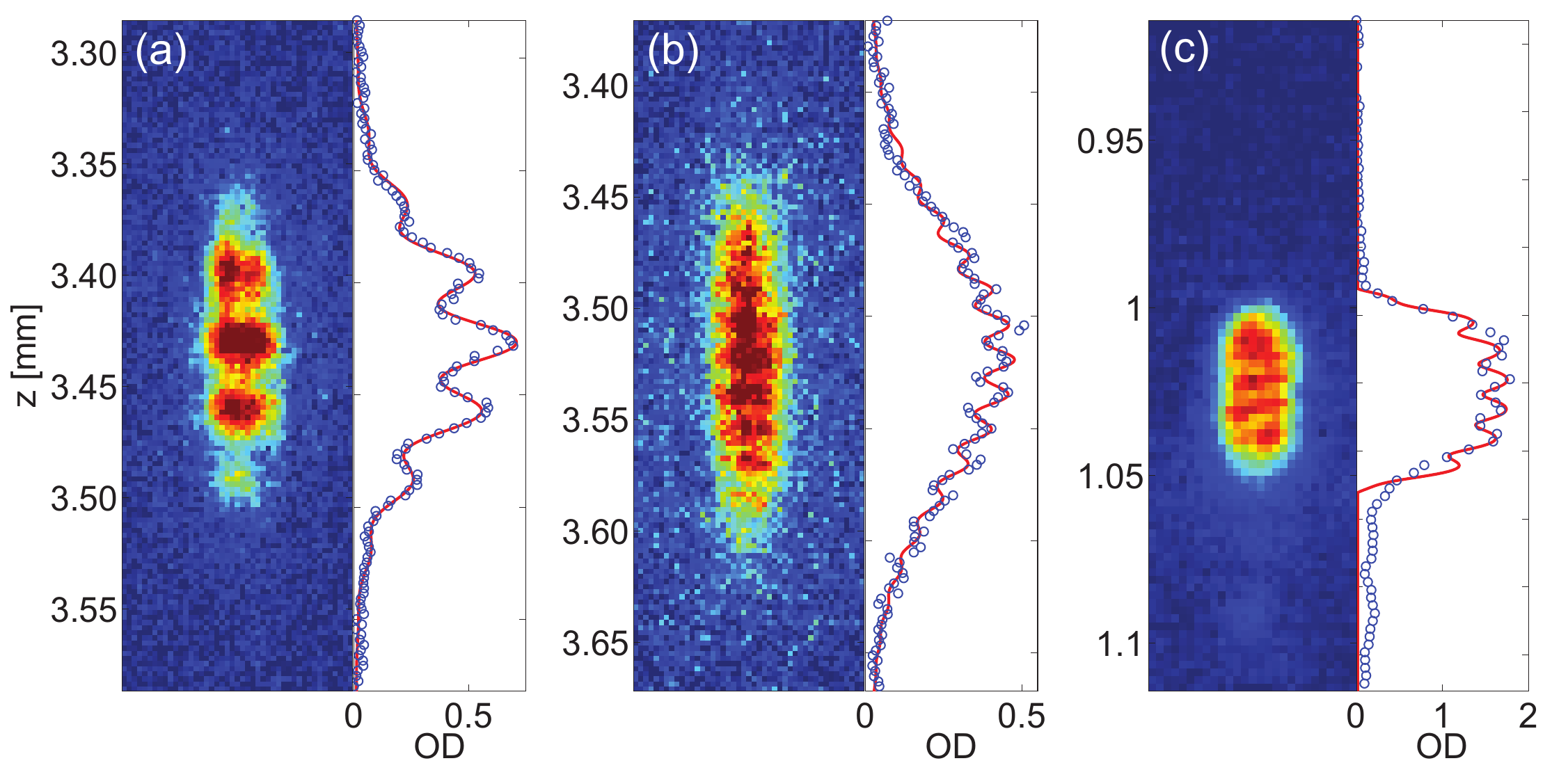}%
\caption[Interference fringes]{
Interference fringes: (a) Large ($T=5\,\mu$s) and (b) small ($T=10\,\mu$s) periodicity fringes, created by recombining the two outputs of the FGBS. Cuts of the optical density (OD) are also provided, with fits (see Methods for fit description).
The two \2 wavepackets were recombined by an additional magnetic gradient and imaged after $14\,$ms of TOF. The different $T$ gives rise to a different $\Delta v$, and hence a different distance before recombination, leading to different fringe spacings, $33\,\mu$m and $16\,\mu$m, respectively.
(c) Fringes observed after the FGBS is applied to a trapped BEC ($T=700\,\mu$s, TOF = $10\,$ms), when recombining the two \2 wavepackets by allowing them to oscillate $2\,$ms in the trap before releasing them. The $8\,\mu$m periodicity fits well with the well known formula for fringe separation $ht/md$.
The low visibility is attributed mostly to our limited optical resolution ($\approx 5\,\mu$m) and our less-than-perfect $\pi/2$ pulses.
}
\label{fig:FGBS:fig6}
\end{figure}

In Fig. \ref{fig:FGBS:fig6}(a,b) we present typical images of interference fringes as observed in our free-fall experiments, where we use an additional gradient pulse at a specific time $\Delta t$ after the first one. At this time, the two wavepackets are already separated by a distance $d=\Delta p\Delta t/m$, $\Delta p$ being the momentum transfer difference due to the FGBS. The second gradient pulse gives a stronger kick to the wavepacket with smaller momentum, which is at this time closer to the chip wire relative to the wavepacket with larger momentum. The duration of the second momentum kick is tuned such that after this kick the two wavepackets have the same momentum (with spatial separation $d$).
They expand and overlap after a TOF to give rise to an interference pattern, which is fitted to
\begin{equation}\label{eq:FGBS:fitG}
A\cdot\exp\left[-\frac{(z-z_0)^2}{2\sigma^2}\right]\left[1+v\sin(\frac{2\pi}{\lambda} (z-z_0)+\phi)\right],
\end{equation}
where $z_0$ and $\sigma$ are the center and length of the joined wavepacket, $\lambda=h t/md$ and $\phi$ are the wavelength and phase of the fringes, and $v$ is the visibility.

Fig. \ref{fig:FGBS:fig6}(c) shows an interference pattern generated with a BEC split in the trap. Here we have allowed the atoms to oscillate in the trap for a period of about $2\,$ms, which is approximately a quarter of the trap harmonic period, so that the relative velocity between the two wavepackets is almost zero, see Fig. \ref{fig:FGBS:fig4}(d). The trap is then released and the two wavepackets, positioned at $z\approx 0$ and at $z\approx \Delta p/m\omega\equiv d$, expand, overlap, and form multiple interference fringes. 
The fit in Fig. \ref{fig:FGBS:fig6}(c) used a form similar to Eq. \eqref{eq:FGBS:fitG} but with a Thomas-Fermi envelope
\begin{equation}\label{eq:FGBS:fitTF}
f \equiv A\cdot \text{max}\{1-\frac{(z-z_0)^{2}}{\text{w}^{2}},0\}^{\frac 3 2}\left[1+v\sin(\frac{2\pi}{\lambda}(z-z_{0})+\phi)\right]
\end{equation}
and the data was actually fitted to the form
\begin{equation}
\frac {f}{1+f/A_{sat}},
\end{equation}
which takes into account a reduction due to saturation.

\section{Phase and momentum stability}

Beyond the presentation and characterization of the technique and kinematics of the FGBS, 
it is important to analyze the prospects and bounds on the stability of the phase and momentum differences, $\Delta \phi$ and $\Delta p$, between the two outports of the FGBS. These stabilities determine the quality of interferometry that the FGBS can support. For a general FGBS,
\begin{equation}
\Delta \phi = (V_2-V_1) T/\hbar~\text{and}~\Delta p = (F_2-F_1) T,
\end{equation}
and the stability is derived from the specific implementation of the FGBS. For magnetic fields generated from thin chip wires and considering only the linear Zeeman effect (with our \1 and \2 states), the FGBS imparts
\begin{eqnarray}
\Delta \phi \propto \frac {IT}{z}~\text{and}~\Delta p \propto \frac {IT}{z^{2}},
\end{eqnarray}
and consequently $\delta\phi/\Delta\phi$ and $\delta p/\Delta p$, the relative instabilities at the output of the FGBS, are proportional to $\delta I /I$, $\delta T/T$ and $\delta z/z$. Our main source of instability, which led to  $\delta\phi\sim 2\pi$, was the latter term, as the trapping position was not determined by the gradient inducing wire but rather by a distant wire.

The position instability can be made negligible by a trapping procedure based on a single current such that the trap position depends only on the geometry and is independent of the current. For example, let us use, for both the trapping and the momentum kick, three parallel chip wires (distance $D$) connected in series, where the two side wires replace the external bias field.
The trap position $z=D$ is independent of $I$.
As $B=\frac{\mu_0 I}{2\pi}\left[\frac{1}{z}-\frac{2z}{z^2+d^2}\right]$ and $\delta B=\frac{\delta I}{I}B+\frac{\partial B}{\partial z}\delta z$, the small $\delta z$ will suppress the second term in $\delta B$. 
In addition, the created quadrupole will
suppress the first term as well, as it has the same gradient as a single wire but with a much weaker magnetic field. Consequently, $\delta \phi$ will be made to depend only on $\delta T$ as $\phi \propto BT$.

In order to estimate the bounds on the $\delta p/\Delta p$ stability, one should consider the available technology. Let us assume a $10\,\mu$s pulse. Then, for a $2\,$A current (containing $\sim10^{14}$ electrons), the shot noise leads to $\delta I /I\sim 10^{-7}$. Power sources with sub-shot noise are being developed, e.g. at JPL \cite{NasaTechBriefs}, and may enable an even better stability (one obviously needs to also account for flicker and Johnson-Nyquist noise). Ultra stable capacitor driven current pulses may also improve performance (at mK temperature stability, available capacitors reach $\delta C /C =10^{-9}$). For pico-second switching electronics, one similarly finds $\delta T/T\sim 10^{-7}$. We note that for longer pulses, the $\delta T/T$ stability will improve linearly with time (e.g. $\sim 10^{-9}$ for a one mili-second pulse), and thus with momentum transfer (as long as $\delta I /I$ is not the limiting factor). Overall, a momentum stability of $10^{-7}$ and beyond, should be considered feasible.

In the trapped BEC experiment, the main source of instability is different. The $g/2$ gradient is fixed and does not fluctuate, and because of the long time between the pulses, $\delta T/T$ can also be neglected. The main limitation is the shot-to-shot fluctuations of the magnetic field at the trap bottom, which changes the energy between the two levels.

\section{Conclusions and outlook}

To conclude, we have presented a new beam-splitting scheme. The FGBS can make use
of different types of potentials and internal states.
The scheme utilizes a Ramsey-like process which introduces a differential acceleration between two wavepackets by a state selective gradient. In the specific realization presented here, the FGBS makes use of magnetic gradients and Zeeman sub-levels. The FGBS is shown to enable a very large splitting momentum in very short times while also allowing for a wide dynamic range. 
The new beam-splitter may be used for freely propagating atoms as well as a trapped BEC and may be utilized for future interferometry experiments both for fundamental studies and technological applications.

This work focused on the spatial splitting of an atomic wavepacket and we have briefly presented an interferometric signal based on spatial fringes. As an outlook, and in order to further emphasize the versatility of the FGBS, let us note that while the present work made use of pulses in the time domain, a continuous beam apparatus may make use of pulses in the spatial domain.
In addition, we consider 
the possibility of the FGBS to also give rise to a signal based on the population of internal states.
Furthermore, we also show that different internal states may be used by the FGBS.

\subsection{Interference with population fringes\\}

Typical interferometers probe the phase between two propagation paths by employing a second beam splitter to translate this phase into populations of momentum states or internal atomic states.
Such an interferometer based on the FGBS would include an appropriate ``mirror" that will bring the two wavepackets back to the same point in space, each with a different momentum state. At this point, we apply another FGBS sequence that would give an appropriate differential momentum kick. The quantum state of an atom just before the operation of the second FGBS is
\begin{equation}
\frac{1}{\sqrt{2}}[|2,\bar{p}+\Delta p/2\rangle+e^{i\phi}|2,\bar{p}-\Delta p/2\rangle],
\end{equation}
where $\bar{p}=\frac{1}{2}(p_1+p_2)$ is the average momentum just before the second FGBS,  $\Delta p$ is the differential momentum due to the first FGBS, and $\phi$ is the phase difference between the two paths which was accumulated during the propagation. Based on Eq. \eqref{eq:FGBS:general}, the state after the second FGBS, which again introduces a differential momentum $\Delta p$, would give
\begin{eqnarray}
&&\frac{e^{i\phi/2}}{\sqrt{2}}[\cos(\phi/2)|1\rangle+\sin(\phi/2)|2\rangle]|\bar{p}_f\rangle+ \\
&& \frac{1}{2\sqrt{2}}[(|1\rangle+|2\rangle)|\bar{p}_f+\Delta p\rangle
    +e^{i\phi}(|1\rangle-|2\rangle)|\bar{p}_f-\Delta p\rangle], \nonumber
\end{eqnarray}
where $\bar{p}_f$ is the average final momentum after the interferometer.
The phase $\phi$ can then be determined by a selective detection of the population of the internal state of the atoms with the momentum state $|\bar{p}_f\rangle$.
The drawback of this interferometry scheme is that half of the atoms are discarded in each beam-splitter and only $1/4$ of the atoms contribute to the interferometric signal.

\subsection{Splitting magnetic insensitive states\\}

Another interferometric scheme which is based on the same idea of the FGBS involves the two magnetically insensitive sublevels $|1,0\rangle$ and  $|2,0\rangle$ of the two hyperfine states, which are analogous to those used in present-day precision interferometers. A micro-wave $\pi/2$ pulse may be used to create an equally populated superposition of the two states and then they may be split into two momentum states using a magnetic gradient at a high magnetic field. The nonlinear Zeeman shift of the transition energy between the states is $\Delta E\approx \alpha B^2$, where $\alpha= 2\pi\hbar\times 575\,$Hz/G$^2$. At a distance of $10\,\mu$m from a wire carrying $2\,$A of current, the atoms are exposed to a magnetic field of $B=400\,$G and a magnetic
gradient of $\partial_z B=40\,$kG/mm. It then follows that $F=\partial_z\Delta E(z)
=2\pi\hbar\times 575\times 2B\partial_z B=1.22\times 10^{-20}\,$N and consequently atoms receive a differential velocity of $84.7\,$mm/s for a $1\,\mu$s pulse, equivalent to about $18\,\hbar k$. As the two states are relatively magnetically insensitive, a second $\pi/2$ pulse would not be needed and the two output beams of this beam splitter could be used for interferometry in a completely analogous way to existing interferometers based on light beam splitters, by, for example, employing a second $\pi/2$ pulse when the wavepackets are recombined, and measuring the populations of the internal states.

\section*{Acknowledgment}
We are most grateful to David Groswasser, Zina Binshtok, Shuyu Zhou, Mark Keil, Daniel Rohrlich and the rest of the AtomChip group for their assistance.

\end{document}